# Measuring large optical reflection matrices of turbid media

Hyeonseung Yu[a], Jung-Hoon Park[a,b], YongKeun Park[a,*]

[a] *Department of Physics, Korea Advanced Institute of Science and Technology, Daejeon 305-701, Republic of Korea*

[b] *Current affiliation: Howard Hughes Medical Institute, Janelia Research Campus, 19700 Helix Drive, Ashburn, Virginia 20147, USA*

[*] *e-mail:* yk.park@kaist.ac.kr

Abstract: We report the measurement of a large optical reflection matrix (RM) of a highly disordered medium. Incident optical fields onto a turbid sample are controlled by a spatial light modulator, and the corresponding fields reflected from the sample are measured using full-field Michelson interferometry. The number of modes in the measured RM is set to exceed the number of resolvable modes in the scattering media. We successfully study the subtle intrinsic correlations in the RM which agrees with the theoretical prediction by random-matrix theory when the effect of the limited numerical aperture on the eigenvalue distribution of the RM is taken into account. The possibility of the enhanced delivery of incident energy into scattering media is also examined from the eigenvalue distribution which promises efficient light therapeutic applications.

## 1. Introduction

Wave propagation through disordered media is a fundamental physical phenomenon, where the input and the scattered wave field are linearly related via the scattering matrix (SM). A SM is formed by two transmission matrices (TM) and two reflection matrices (RM) taking into account the two sides of the medium which can act both as the input or output faces. Recently TMs of turbid medium have been optically measured [1-9], and various utilizations of TMs have been demonstrated including the reconstruction of an original image from a speckle field that has transmitted through a turbid layer [10], optical manipulation [11], depth-enhanced wavefront shaping optical coherence tomography [12, 13], the enhancement of energy delivery through scattering media [14, 15], the study of energy deposition in complex media [16], and the subwavelength focusing and imaging using randomly distributed nanoparticles [17, 18]. Furthermore, TMs can also provide important information for wavefront shaping techniques [19-25].

However, considering that these approaches have potentials for imaging or light delivery through biological tissues [26], the key breakthrough for practical applications may lie in the utilization of RMs of scattering media since optical instruments adequate for *in vivo* biomedical applications must function in reflection geometry. Recently, an acoustic RM of turbid media has been measured [27] and it is demonstrated that the measurement of full scattering matrix including both the RM and TM can address the perfect transmission channel [28]. However, experimental investigations on the properties of optical RMs of turbid media have not been fully conducted. In optics, a RM has been first measured for weakly scattering media [29], which can be regarded as an aberration layer. More recently, a RM of complex media has been measured using interferometric method, and the suppression of reflected light intensity has been demonstrated using the measured RM [30]. The method to measure the time-resolved RM is also proposed, which provides information on the light field reflected from the specific depth via low coherence interferometry. It enables the enhanced light delivery at the target depth inside the scattering media [31].

Here we present the measurement of a large optical RM of scattering media and address the statistical property of measured RM. The input wave fields are modulated by a spatial light modulator (SLM) and the corresponding reflected fields from the sample are measured by a full-field Michelson interferometer. The optical modes are sampled finely so that the measured total mode number exceeds the total number of resolvable modes for the field-of-view (FoV) in the sample. After constructing the RM of the sample, we investigate the eigenvalue distribution of the acquired RM and made comparisons with the theoretical prediction by random-matrix theory (RMT). Recent theoretical [32] and experimental studies [3] have shown that the acquirable information from SMs is limited by the numerical aperture (NA) of the optical imaging system. We study this limitation on the RM through a comparison between our simulation model and experimental results.

## 2. Experimental Setup

The experimental setup is depicted in Fig. 1. The setup comprises of a SLM and a Michelson interferometer to modulate and measure the respective optical light fields. A linearly-polarized collimated light beam from a HeNe laser (HNL020R, Thorlabs Inc.) is divided into a sample and a reference beam at a beam splitter. The beam impinged onto a scattering sample is phase-modulated by a SLM (LCOS-SLM X10468-02, Hamamatsu Photonics, Japan). A half-wave plate is used to rotate the polarization of the reference beam. The reflected beam from a scattering sample is projected onto a CCD camera (IMI Tech., IMC-30FC, Republic of Korea), and interferes with the reference beam to form an off-axis hologram, from which the amplitude and phase of the reflected field are retrieved via the Hilbert transform [33].

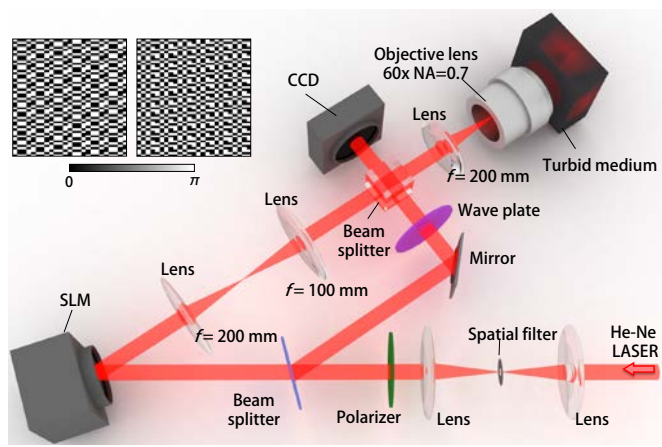

**Fig. 1.** Schematic of the experimental setup. The inset shows representative Hadamard patterns for phase modulation.

Since the field acquisition requires only a single-shot measurement of an off-axis hologram, the acquisition speed is limited by the recording speed of the detector and the modulation speed of input fields. In the current setup, the major limiting factor of the acquisition time is the refresh rate of the SLM which is 30 ms. To avoid cross-talk between successive measurements, the measurement time for a reflected field was set as 50 ms, resulting in the total measurement time of approximately 10 min for 12,288 incident modulated fields.

In order to systematically and effectively modulate incident light fields, we employed the Hadamard patterns as the input basis. The amplitude of the modulated field is uniform, and the phase map, having the values of either 0 or $\pi$, is generated from the Hadamard patterns. The representative phase maps are shown in the inset of Fig. 1. The lateral magnification of the optical system is designed such that, when the phase patterns are projected onto a scattering sample, the minimum size of the phase patterns corresponds to the diffraction-limited spot size at the sample plane. In principle, these phase-modulated light fields span the same information that can be assessed by the light fields scanned in angular spectrum space, i.e. by tilting the illumination angle of the incident beam via a galvanometer mirror. Although deploying plane-waves with tilting illumination angles using a galvanometer-mirror may provide high-speed measurements as demonstrated in the measurements of TMs [1, 14], the use of a SLM has the following advantages: (a) no mechanical parts eliminates mechanical noise; (b) a SLM can also be used for applying any arbitrary shaped wavefront as well as measuring RMs of a turbid medium.

The scattering sample used for the experiments is a layer of ZnO nanoparticles (mean diameter = 200 nm) with a thickness of $160 \pm 15$ μm. Using an integrating sphere, the mean-free path of the sample was measured as $1.05 \pm 0.11$ μm. The FoV of the measurement was 27.3 μm × 20.5 μm. In 2-D slab waveguide geometry, the number of propagating modes is given as $N = p\pi A/\lambda^2$ [34], where $A$ is the area of a waveguide and $p = 1$ or 2 with respect to the number of measured orthogonal polarization states. In our case, $p = 1$ since only one polarization state is modulated and measured. The total resolved number of modes inside the FoV is then calculated to be 4,381. For experiments, the number of input modes was set to be $M = 12,288$ to oversample all resolvable optical modes.

## 3. Measurement and calibration of RM

For calibration purposes, we first measure the RM of a mirror. When illuminated with a normal incident plane wave, the measured amplitude and phase maps of the light reflected from the mirror show uniform values (Figs.

2a-b). Here the amplitude is normalized so that both the total input and total reflected energy are equal to unity. Then we measure the RM of a scattering medium. Both the measured amplitude and phase maps of the beam reflected from the scattering sample, illuminated with a normal incident plane wave, exhibit highly disordered speckled patterns (Figs. 2c-d). The output intensity from the scattering sample is normalized to the reference data obtained from a mirror.

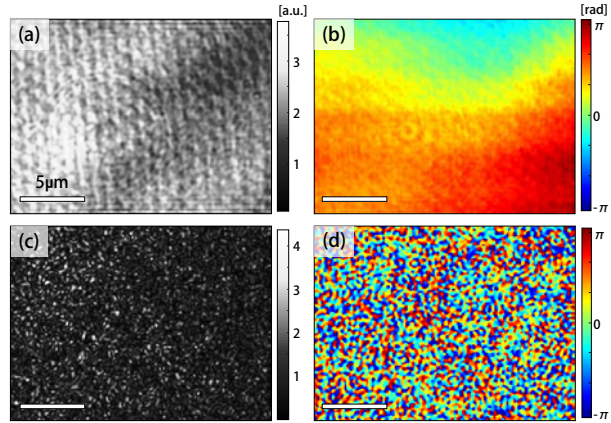

**Fig. 2.** (a) The measured amplitude and (b) phase of a reflected beam from a mirror. (c) The measured amplitude and (d) phase of a reflected beam from a scattering sample.

After acquiring the reflected fields for each input basis, the RM of the mirror is constructed as shown in Fig 3a. The RM of the mirror plays the important role in the analysis since we presumably know that the RM is equal to an identity matrix multiplied by constant phase factor $e^{i\pi}$; the amplitude of the reflected field is the same with that of the input field and the phase is delayed by $\pi$ according to the law of reflection. Then any deviation from this ideal value can be regarded to originate from measurement error. Since the input basis is the Hadamard basis and the output basis is represented in the spatial coordinate basis, the RM of the mirror is expressed in the following way,

$$R_{mirror,ideal} = R_{mirror,exp} KH^{-1}$$
$$R_{mirror,exp} = [H_{1,exp}, \quad H_{2,exp}, \quad \ldots \quad H_{n,exp}]$$
$$H = [H_{1,ideal}, \quad H_{2,ideal}, \quad \ldots \quad H_{n,ideal}]$$

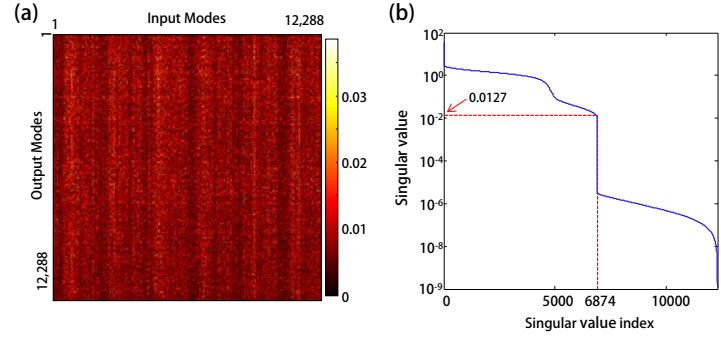

**Fig. 3.** (a) The amplitude of the RM ($R_{mirror,\exp}$) of the mirror. (b) The singular value distribution of the RM of the mirror.

$R_{mirror,ideal}$ is an $M\times M$ identity matrix. $H$ is an $M\times M$ Hadamard transformation and $H_i$ denotes the $i$-th $M\times 1$ Hadamard input vector. $K$ is an $M\times M$ matrix representing the correction to the imperfect measurement. If we assume that the reflection from the mirror is perfect, the calibration term $KH^{-1}$ is purely dependent on the experimental setup and preserved in the measurement of the RM of the scattering sample, providing the following relation.

$$R_{sct} = R_{sct,\exp} KH^{-1}$$

As easily noticeable, the matrix $KH^{-1}$ is the inverse matrix of $R_{mirror,\exp}$. In our case, however, the calculation of the inverse matrix of $R_{mirror,\exp}$ needs careful consideration. The number of sampled modes exceeds the number of the propagating modes, so $R_{mirror,\exp}$ has a rank smaller than the dimension of the matrix. In other words, $R_{mirror,\exp}$ is a singular matrix and does not have an inverse matrix. Due to this reason, we have introduced another correction term $W$ instead of the exact inverse matrix, following the method in Ref. [8]. $W$ is basically an intermediate parameter between the pseudo inverse matrix and the phase conjugation matrix of $R_{mirror,\exp}$, defined as

$$W = [R^{\dagger}_{\exp} \cdot R_{\exp} + \sigma \cdot I]^{-1} \cdot R^{\dagger}_{\exp}$$

Here we omitted the subscript 'mirror' for simplicity. $\sigma$ is the experimental noise level and $I$ is the $M\times M$ identity matrix. Then $W$ satisfies following relations,

$$R_{mirror} = R_{mirror,exp} W$$
$$R_{sct} = R_{sct,exp} W$$

To establish the proper criteria for setting $\sigma$, we first analyzed the singular values of $R_{mirror,exp}$, as plotted in Figure 3b. $R_{mirror,exp}$ has a rank of 6,874, which is 56% larger than the expected value of 4,381. This deviation is possibly due to the sample geometry; the number of propagating modes is calculated for a quasi-2-D wave-guide configuration, however, the measured scattering sample has a 3-D open-slab geometry, which will be discussed in-depth later. As shown in the figure, the singular values of the indices exceeding the rank are close to zero and the smallest singular value within the rank is 0.0127. Near-zero singular values beyond the rank in the original matrix produce abnormally large singular values in the inverse matrix. To prevent this unwanted error, the minimum singular value must be settled. This is accomplished by $\sigma \cdot I$ which adds an additional singular value background to $R^{\dagger}_{exp} \cdot R_{exp}$. Therefore we set $\sigma$ to the smallest singular value 0.0127 corresponding to the singular value index equal to the rank.

The amplitude of the RM of the mirror after calibration with $W$ is shown in Fig. 4a. Although it slightly deviates from the perfect identity matrix, the highest values are concentrated along the diagonal line, demonstrating the validity and sensitivity of the present method. Based on this validation, we use the same approach to reconstruct and calibrate the RM of the scattering sample; Figs. 4b-c respectively show the amplitude and phase of the RM, in which the values of the RM are randomly distributed, clearly demonstrating the effect of multiple light scattering.

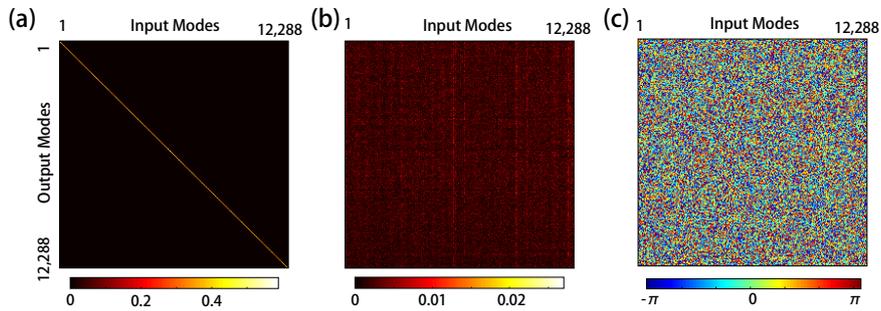

**Fig. 4.** (a) The amplitude of the calibrated RM of the mirror. (b) The amplitude and (c) phase of the calibrated RM of the scattering sample.

## 4. Eigenvalue distribution of RM

To study the statistical properties of the acquired RM of scattering media, we investigate the eigenvalue distribution of the RM and compare with RMT. RMT provides statistical approaches for studying complex structures including neutron resonances in nuclear reactions, quantum chaos, and wave transport through disordered media [35]. A singular vector of a RM can be interpreted as an input eigenchannel, and the square of its corresponding singular value is the reflectivity of that input eigenchannel, which is equal to the eigenvalue $R$ of $R_{ideal} \cdot R^{\dagger}_{ideal}$. Assuming no absorption in the sample, the summation of the reflectivity $R$ and the transmission $T$ is equal to 1.

In RMT, DMPK (Dorokhov-Mello-Pereyra-Kumar) model predicted that the probability density function of a RM of scattering media, exhibiting correlation in the eigenvalues, is given as [36],

$$p(R) = \frac{\langle R \rangle}{2(1-R)\sqrt{R}}, \qquad (2)$$

where $\langle R \rangle$ denotes the mean reflectivity of the distribution. For a random matrix with uncorrelated entries, however, the probability density distribution of its singular values follows the quarter-circle law [37]. As demonstrated in acoustic wave measurements, a RM of turbid media follows the quarter-circle law when the correlation is forcibly eliminated [27]. The distribution simulated using the DMPK model (the black solid line in Fig. 4) show two noticeable characteristics; the existence of open eigenchannels, which transmit almost all incident energy through a turbid layer ($T \approx 1$ and thus $R \approx 0$), and closed eigenchannels, which reflect almost all incident energy ($T \approx 0$ and $R \approx 1$) [35].

In order to experimentally access these intriguing optical modes, the direct and precise measurement of a large RM is crucial. Once the full RM of a turbid medium is measured, the perfect open or closed channels can be accessed in principle. In an experiment, however, the full RM measurement and thus the *direct* observation of the open and closed modes are limited by the optical system, mainly due to the finite NA of the objective lens that illuminates the incident optical fields to the sample and collects the scattered field from the sample.

To analytically study the dependence of the NA on the statistical properties of the RM, we employed a fraction parameter $f$, which measures the fraction of accessible modes for both the illumination and detection sides ($0 < f \leq 1$) [3]. The fraction parameter $f$ can be related to the NA of an optical imaging system as $f = NA^2/n^2$ [1], where

$n$ is the refractive index of the surrounding medium of the sample. The simulated results according to different $f$ values are plotted in Fig. 5, from which several important observations can be made. The open channels ($R \approx 0$) and perfect reflection ($R \approx 1$) are both observed in the RMs with high $f$ values. For RMs with $f$ values smaller than 0.5, the probability to observe perfect reflection channels become zero.

In contrast, the open channels ($R \approx 0$) are always observed in the simulations, and can be experimentally accessible independent of the *NA* of an optical system. However, this interpretation on the open channels is rather misleading. It should be noted that by the definition of the fraction parameter $f$, the value of $R$ means the reflectance from a system consisting of an objective lens and a turbid medium, and not directly from the turbid medium. It is then possible to observe optical modes with $R \approx 0$, but this does not correspond to the open channels, because the light can be reflected without being measured, i.e. outside the NA of the objective or in the orthogonal polarization. Therefore, when $f < 1$, $R = 0$ does not necessarily mean the perfect antireflection; reflected waves that cannot be collected by the objective lens may still exist. Thus, the seemingly increasing probability of observing open channels as $f$ decreases is an artifact due to the light being scattered and not being recollected in the smaller NA.

The eigenvalue distribution calculated from the measured RM is shown in Fig. 5. The $f$ value for the current experimental condition is 0.25 ($n = 1$ and $NA = 0.7$, $p = 1$) and the eigenvalue distribution from the measured RM is indicated by the gray area. The measured and simulated eigenvalue distributions are well matching for low values ($R < 0.1$). However, beyond this threshold, the curves deviate strongly from one another.

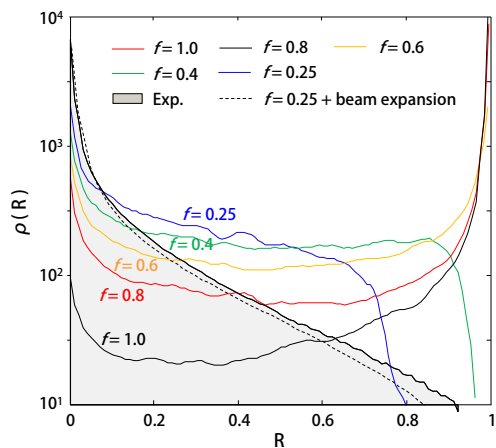

**Fig. 5.** The probability distributions of the eigenvalues of the RM of a turbid media. The experimental result is represented with a gray area. The numerically calculated distributions for various $f$ values are represented with colored solid lines. The dashed-line denotes the simulated result considering the effect of the beam expansion.

To explain this deviation, we should note that the theoretical model assumes a quasi-2-D waveguide configuration, whereas the measured scattering sample has 3-D open-slab geometry. To take into account the energy loss due to the slab geometry, we considered the effect of beam expansion during light transport in the sample. In the present experimental condition, the size of the FoV is the same for both incident and reflected beams, and thus the lateral size expansion of the propagating beam can be regarded as energy loss due to the slab geometry.

To consider the effect of beam expansion, we first simulated a RM for a rectangular waveguide with mode number $N = 4,096$ [38]. The scattering layer is simulated as 300 successive aberrating layers, and the input and output face dimension is $64a \times 64a$, where $a$ is the size of diffraction limited spot at a given wavelength. Applying input beams from the Hadamard basis, the back-scattered field can be obtained from the simulated RM. To take into account the loss due to the slab geometry, only $M = 2,190$ components of each input field vectors have non-zero values and the rest are filled with zeros. This accounts for the 87% size expansion for the reflected beams compared to the incident beam, where the expansion value was experimentally confirmed. Then, we took the partial $M \times M$ submatrix from the RM to reflect the loss due to the slab geometry. Finally, the fraction parameter $f$ is applied to the reduced RM, and its resulting eigenvalue distribution is represented in a black-dashed line in Fig. 5. The resulting RM simulation shows reasonable agreement with the measured eigenvalue distribution over a large range of $R$. The maximum eigenvalue of the measured RM is slightly higher than that of the theoretical RM, which can be explained by imperfect choice of $\sigma$ in $W$.

## 5. Conclusion

We demonstrated the measurement of a large optical RM of scattering media. Using a Michelson interferometer equipped with a SLM, the incident optical fields were systematically modulated and the corresponding back-scattered fields from the scattering medium were precisely measured, from which the RM was constructed. The measured RM has a size of $12,288 \times 12,288$, which exceeds the number of total resolvable modes in the FoV (27.3 μm × 20.5 μm) of the turbid sample. The eigenvalue distribution of the measured RM was obtained and agreement with the RMT theory was verified, revealing the intrinsic correlations in the RM.

Our measurement of the large RM of turbid media implies interesting consequences. The limited NA in an optical imaging system acts as an insurmountable barrier to measuring the complete RM, and thus the closed channel or perfect reflection ($R = 1$) is technically inaccessible with the partial measurement of RM. Though optical modes with the eigenvalues of 0 are still observed in the partial measurement of RM obtained with limited NA, however, physical meaning of these optical eigenmodes should be carefully interpreted; the optical modes with the eigenvalues of 0 or 1 in measured RM are not 'perfect' closed or open eigenchannels in the turbid media. A reflectivity from a specific channel $i$, $R_i$ can be precisely expressed as $R_i = 1 - T_i - R_c$, where $T_i$ is the transmittance and $R_c$ is the reflectivity for a reflected beam which is not collectable with limited NA. For $f < 1$, $R_c$ cannot be determined since the measured RM does not contain any information on $R_c$. Thus the perfect energy transmission is only guaranteed when $f = 1$. Consequently, experimentally accessing both the 'perfect' open and closed channels will be challenging in practical imaging systems employing limited *NA*.

The method presented here provides an effective tool enabling the study of subtle intrinsic correlations of the back-scattering characteristics of a disordered system. We expect this study may open new possibilities in experimental and theoretical studies of multiple light scattering in complex systems.

**Acknowledgements**

The authors thank Young-Jin Kim, Seongheum Han, Seung-Woo Kim (Department of Physics, KAIST), and Ji Oon Lee (Department of Mathematical Science, KAIST) for providing optical components and for helpful discussions. This work was funded by KAIST - Khalifar University Project, APCTP, MEST/NRF [2012R1A1A1009082, 2014K1A3A1A09063027, 2013R1A1A3011886, 2012-M3C1A1-048860, 2013M3C1A3063046, 2014M3C1A3052537].